\documentclass{ws-ijmpd}
\usepackage[super,compress]{cite}

\usepackage{hyperref,amsfonts,amsmath,amssymb,graphicx,bm,subfigure,cite}

\usepackage[T1]{fontenc}

\numberwithin{equation}{section}

\begin{document}

\markboth{Maxim Dvornikov}
{Scattering of neutrinos by black hole accounting for interaction with accretion disk}

%
\catchline{}{}{}{}{}
%

\title{SCATTERING OF NEUTRINOS BY A ROTATING BLACK HOLE ACCOUNTING FOR THE
ELECTROWEAK INTERACTION WITH AN ACCRETION DISK}

\author{MAXIM DVORNIKOV}

\address{Pushkov Institute of Terrestrial Magnetism, Ionosphere \\ and Radiowave Propagation (IZMIRAN), \\
108840 Moscow, Troitsk, Russia\\
maxdvo@izmiran.ru}

\maketitle

\begin{abstract}
We study spin effects in the neutrino gravitational scattering by
a supermassive black hole with a magnetized accretion disk having
a finite thickness. We exactly describe the propagation of ultrarelativistic
neutrinos on null geodesics and solve the spin precession equation
along each neutrino trajectory. The interaction of neutrinos with
the magnetic field is owing to the nonzero diagonal magnetic moment.
Additionally, neutrinos interact with plasma of the accretion disk electroweakly
within the Fermi approximation. These interactions are obtained to
change the polarization of incoming neutrinos, which are left particles.
The fluxes of scattered neutrinos, proportional to the survival probability
of spin oscillations, are derived for various parameters of the system.
In particular, we are focused on the matter influence on the outgoing
neutrinos flux. The possibility to observe the predicted effects for
astrophysical neutrinos is briefly discussed.
\end{abstract}

\section{Introduction}

Neutrino interactions with external fields can significantly modify
the dynamics of neutrino oscillations, which were recently confirmed
experimentally (see, e.g., Ref.~\refcite{Ace22}). We deal here mainly
with neutrino spin oscillations, in which active left handed neutrinos
become sterile right particles under the influence of some external
fields. Gravitational interaction, in spite of its weakness, can also
contribute to neutrino oscillations (see, e.g., Ref.~\refcite{PirRoyWud96}).

The propagation of a spinning particle in curved spacetime was first
considered in Ref.~\refcite{Pap51}. Then, the quasiclassical approach
for the description of the spin evolution of a point-like elementary
particle in a gravitational field was formulated in Ref.~\refcite{PomKhr98}.
This method was applied for the studies of neutrino spin oscillations
in a curved spacetime both in frames of the General Relativity (GR)
in Refs.~\refcite{AlaNod13,BayPen21} and in various alternative gravity
theories in Refs.~\refcite{AlaNod15,Cha15,MasLam21}. The perturbative
quantum treatment of neutrino spin oscillations in the gravitational
field of a black hole (BH) was carried out in Refs.~\refcite{SorZil07,Sor12}.
Quasiclassical and quantum approaches for the studies of spinning
particles in a gravitational field were reviewed in Ref.~\refcite{Ver22}.

We considered neutrino spin oscillations in curved spacetimes in Refs.~\refcite{Dvo06,Dvo13,Dvo20a,Dvo20,Dvo21,Dvo23,Dvo19}.
Both stationary~\cite{Dvo06,Dvo13,Dvo20a,Dvo20,Dvo21,Dvo23} and
time dependent metrics, like a gravitational wave~\cite{Dvo19},
were analyzed. We studied spin effects in the neutrino gravitational
scattering by BH in Refs.~\refcite{Dvo20a,Dvo20,Dvo21,Dvo23}. In a
scattering problem, both in- and out-states are in the asymptotically
flat spacetime. Thus, we can attribute certain polarization to these
states. We studied the cases of both nonrotating and rotating BHs,
as well as analyzed the contribution of various external fields to
spin oscillations of scattered neutrinos. In the present work, we
continue to examine this problem.

The recent observations of the shadows of supermassive BHs (SMBHs)
in the centers of M87 and our Galaxy by the Event Horizon Telescope
(EHT)~\cite{Aki19,Aki22} were the motivation for the present research.
Those results are the unique tests of GR in the strong field limit.
There are searches of the high energy neutrinos emission from the
centers of some active galaxies~\cite{Abb22}. If such neutrinos
are produced in the vicinity of the SMBH surface, they are strongly
lensed analogously to photons observed by the EHT collaboration. External
fields, which can exist near SMBHs, will change the neutrino polarization
leading to the distortion of the outgoing neutrinos flux. Alternatively,
neutrinos emitted in a core-collapsing supernova (SN) explosion can
be gravitationally lensed by the SMBH in the center of our Galaxy.
Such possibility was discussed in Refs.~\refcite{MenMocQui06,Los21}.
In this situation, we can also expect the influence of external fields
in the curved spacetime on the dynamics of neutrino spin oscillations.

This work is organized in the following way. We start in Sec.~\ref{sec:MOTION}
with a brief description of the ultrarelativistic neutrinos motion in the Kerr metric. We also describe the neutrino spin evolution in
background matter under the influence of electromagnetic and gravitational
fields. Then, in Sec.~\ref{sec:MATTER}, we set up the neutrino interaction
with plasma of an accretion disk. The rest of the parameters of the
system is chosen in Sec.~\ref{sec:PARAM}. We present the results
for the calculation of the outgoing neutrino fluxes, which account
for neutrino spin oscillations, in Sec.~\ref{sec:RES}. The conclusion
is given in Sec.~\ref{sec:CONCL} We find the component of the four
velocity of plasma in the accretion disk in~\ref{sec:UtDERIV}.
Some computational details are provided in~\ref{sec:ADAMS}.

\section{Motion of a neutrino in the Kerr metric and its spin evolution\label{sec:MOTION}}

We study the propagation of a test neutrino in the gravitational field
of a rotating BH which is described by the Kerr metric,
\begin{equation}
\mathrm{d}s^{2}=g_{\mu\nu}\mathrm{d}x^{\mu}\mathrm{d}x^{\nu}=\left(1-\frac{rr_{g}}{\Sigma}\right)\mathrm{d}t^{2}+2\frac{rr_{g}a\sin^{2}\theta}{\Sigma}\mathrm{d}t\mathrm{d}\phi-\frac{\Sigma}{\Delta}\mathrm{d}r^{2}-\Sigma\mathrm{d}\theta^{2}-\frac{\Xi}{\Sigma}\sin^{2}\theta\mathrm{d}\phi^{2},\label{eq:Kerrmetr}
\end{equation}
where the Boyer-Lindquist coordinates $x^{\mu}=(t,r,\theta,\phi)$
are utilized. In Eq.~(\ref{eq:Kerrmetr}), we use the following notations:
\begin{equation}
\Delta=r^{2}-rr_{g}+a^{2},\quad\Sigma=r^{2}+a^{2}\cos^{2}\theta,\quad\Xi=\left(r^{2}+a^{2}\right)\Sigma+rr_{g}a^{2}\sin^{2}\theta.\label{eq:dsxi}
\end{equation}
The mass of BH in Eqs.(\ref{eq:Kerrmetr}) and~(\ref{eq:dsxi}) is
$M=r_{g}/2$ and its angular momentum is $J=Ma$, where $0<a<M$.
The BH spin is directed upward from the equatorial plane $\theta=\pi/2$.

The arbitrary motion of a test ultrarelativistic particle in the Kerr
metric is characterized by two conserved quantities: the angular momentum
$L$ and the Carter constant $Q$. If we study the scattering problem,
$Q>0$. The law of motion and the form of a trajectory for such a
particle can be found in quadratures. The corresponding quantities
are expressed in terms of elliptic integrals. The detailed description
of this problem can be found, e.g., in Ref.~\refcite{Cha83}.

If a test particle is spinning, like a neutrino, the invariant equation
for the evolution of the particle spin $S^{\mu}$ in a curved spacetime
under the influence of the electromagnetic field $F_{\mu\nu}$ and
the background matter has the form~\cite{Dvo13},
\begin{align}
\frac{\mathrm{D}S^{\mu}}{\mathrm{d}\tau}= & 2\mu\left(F^{\mu\nu}S_{\nu}-U^{\mu}U_{\nu}F^{\nu\lambda}S_{\lambda}\right)+\sqrt{2}G_{\mathrm{F}}E^{\mu\nu\lambda\rho}G_{\nu}U_{\lambda}S_{\rho},\label{eq:BMTcurvedst}
\end{align}
where $U^{\mu}=\tfrac{\mathrm{d}x^{\mu}}{\mathrm{d}\tau}$ is the
four velocity in the world coordinates, $\tau$ is the proper time,
$\mathrm{D}S^{\mu}$ is the covariant differential, $E^{\mu\nu\lambda\rho}=\tfrac{1}{\sqrt{-g}}\varepsilon^{\mu\nu\lambda\rho}$
is the covariant antisymmetric tensor in a curved spacetime, $g=\det(g_{\mu\nu})$
is the determinant of the metric tensor, $\mu$ is the neutrino magnetic
moment, $G_{\mathrm{F}}=1.17\times10^{-5}\,\text{GeV}^{-2}$ is the
Fermi constant, and $G_{\mu}$ is the vector which incorporates the
characteristics of the background matter. We discuss its form in details
in Sec.~\ref{sec:MATTER} below. Although Eq.~(\ref{eq:BMTcurvedst})
is written down for massive particles, it has the appropriate limit
for an ultrarelativistic neutrino. In Eq.~(\ref{eq:BMTcurvedst}),
we suppose that a neutrino is a Dirac particle. Despite some claims
about the Majorana nature of neutrinos, this issue is still unresolved~\cite{Bil20}.

It is more convenient to describe the evolution of the neutrino polarization
in a locally Minkowskian frame $x_{a}=e_{a}^{\ \mu}x_{\mu}$, where
$e_{a}^{\ \mu}$ are the vierbein vectors which have the form,
\begin{align}
e_{0}^{\ \mu}= & \left(\sqrt{\frac{\Xi}{\Sigma\Delta}},0,0,\frac{arr_{g}}{\sqrt{\Delta\Sigma\Xi}}\right),\quad e_{1}^{\ \mu}=\left(0,\sqrt{\frac{\Delta}{\Sigma}},0,0\right),\nonumber \\
e_{2}^{\ \mu}= & \left(0,0,\frac{1}{\sqrt{\Sigma}},0\right),\quad e_{3}^{\ \mu}=\left(0,0,0,\frac{1}{\sin\theta}\sqrt{\frac{\Sigma}{\Xi}}\right).\label{eq:vierbKerr}
\end{align}
These vectors diagonalize the metric in Eq.~(\ref{eq:Kerrmetr}),
$g_{\mu\nu}e_{a}^{\ \mu}e_{b}^{\ \nu}=\eta_{ab}$, where $\eta_{ab}=(1,-1,-1,-1)$
is the Minkowski metric tensor. If we define the three vector of the
polarization $\bm{\zeta}$ in the particle rest frame in the locally
Minkowskian frame, it obeys the equation,
\begin{equation}
\frac{\mathrm{d}\bm{\bm{\zeta}}}{\mathrm{d}t}=2(\bm{\bm{\zeta}}\times\bm{\bm{\Omega}}),\quad\bm{\bm{\Omega}}=\bm{\bm{\Omega}}_{g}+\bm{\bm{\Omega}}_{\mathrm{em}}+\bm{\bm{\Omega}}_{\mathrm{matt}}\label{eq:nuspinrot}
\end{equation}
where
\begin{align}\label{eq:vectG}
  \bm{\bm{\Omega}}_{g} & =\frac{1}{2U^{t}}
  \left[
    \mathbf{b}_{g}+\frac{1}{1+u^{0}}
    \left(
      \mathbf{e}_{g}\times\mathbf{u}
    \right)
  \right],
  \nonumber
  \\
  \bm{\bm{\Omega}}_{\mathrm{em}} & =\frac{\mu}{U^{t}}
  \left[
    u^{0}\mathbf{b}-\frac{\mathbf{u}(\mathbf{u}\mathbf{b})}{1+u^{0}}+(\mathbf{e}\times\mathbf{u})
  \right],
  \nonumber
  \\
  \bm{\bm{\Omega}}_{\mathrm{matt}} & =\frac{G_{\mathrm{F}}}{\sqrt{2}U^{t}}
  \left[
    \mathbf{u}
    \left(
      g^{0}-\frac{(\mathbf{gu})}{1+u^{0}}
    \right)-\mathbf{g}
  \right].
\end{align}
Here $u^{a}=(u^{0},\mathbf{u})=e_{\ \mu}^{a}U^{\mu}$, $G_{ab}=(\mathbf{e}_{g},\mathbf{b}_{g})=\gamma_{abc}u^{c}$,
$\gamma_{abc}=\eta_{ad}e_{\ \mu;\nu}^{d}e_{b}^{\ \mu}e_{c}^{\ \nu}$
are the Ricci rotation coefficients, the semicolon stays for the covariant
derivative, $f_{ab}=e_{a}^{\ \mu}e_{b}^{\ \nu}F_{\mu\nu}=(\mathbf{e},\mathbf{b})$
is the electromagnetic field tensor in the locally Minkowskian frame,
and $g^{a}=(g^{0},\mathbf{g})=e_{\ \mu}^{a}G^{\mu}$. The details
of the derivation of Eqs.~(\ref{eq:nuspinrot}) and~(\ref{eq:vectG})
can be found in Ref.~\refcite{Dvo13}.

Instead of dealing with Eq.~(\ref{eq:nuspinrot}), we can rewrite it 
in the form of the effective Schr\"{o}dinger equation
$\mathrm{i}\dot{\psi}=H\psi$, where $H=-(\bm{\sigma}\bm{\Omega})$
and $\bm{\sigma}$ are the Pauli matrices. It is more convenient to
label a point on a neutrino trajectory with $r$ rather than with
$t$. Thus, we transform the Schr\"{o}dinger equation to the form,
\begin{equation}\label{eq:Schrod}
  \mathrm{i}\frac{\mathrm{d}\psi}{\mathrm{d}r}=H_{r}\psi,
  \quad
  H_{r}=-\frac{\mathrm{d}t}{\mathrm{d}r}\mathcal{U}_{2}(\bm{\sigma\Omega})\mathcal{U}_{2}^{\dagger},
\end{equation}
where the matrix $\mathcal{U}_{2}=\exp(\mathrm{i}\sigma_{2}\pi/4)$
accounts for the fact that incoming and outgoing neutrinos move oppositely
and along the first axis in the locally Minkowskian frame. The derivative
$\mathrm{d}t/\mathrm{d}r$ is taken on the basis of the law of motion of neutrinos;
cf., e.g., Refs.~\refcite{Dvo23,Cha83}.

The initial spin wavefunction is $\psi_{-\infty}^{\mathrm{T}}=(1,0)$.
It corresponds to a left polarized neutrino. After reconstructing
the neutrino trajectory and solving Eq.~(\ref{eq:Schrod}), we get
the wave function $\psi_{+\infty}^{\mathrm{T}}=(\psi_{+\infty}^{(\mathrm{R})},\psi_{+\infty}^{(\mathrm{L})})$ of an outgoing neutrino.
The survival probability for a neutrino to remain left polarized after
the scattering is $P_{\mathrm{LL}}=\left|\psi_{+\infty}^{(\mathrm{L})}\right|^{2}$.
Here we account for the fact that the neutrino velocity changes its
direction in the locally Minkowskian frame.

\section{Neutrino interaction with background matter\label{sec:MATTER}}

In this section, we discuss in details the electroweak interaction
of scattering neutrinos with plasma of an accretion disk around BH.
The properties of the accretion disk are also considered.

We suppose that the accretion disk consists of the hydrogen plasma.
The neutrino interaction with electrons and protons is treated within
the Fermi model in the forward scattering approximation. The effective
Lagrangian for the interaction of the neutrino bispinor $\nu$ with
matter has the form,
\begin{equation}
\mathcal{L}_{\mathrm{matt}}=-\frac{G_{\mathrm{F}}}{\sqrt{2}}\bar{\nu}\gamma^{\mu}(1-\gamma^{5})\nu\cdot G_{\mu},
\end{equation}
where $\gamma^{\mu}$ and $\gamma^{5}$ are the Dirac matrices. The
matter characterictics are in the four vector potential $G^{\mu}$
which has the form~\cite{Dvo13},
\begin{equation}
G^{\mu}=\sum_{f=e,p}\left(q_{f}^{(1)}J_{f}^{\mu}+q_{f}^{(2)}\Lambda_{f}^{\mu}\right),\label{eq:Gmu}
\end{equation}
where $J_{f}^{\mu}$ is the invariant hydrodynamics current of plasma
fermions, $\Lambda_{f}^{\mu}$ is the plasma invariant polarization,
and~\cite{DvoStu02}
\begin{equation}
q_{f}^{(1)}=I_{\mathrm{L}3}^{(f)}-2Q_{f}\sin^{2}\theta_{\mathrm{W}}+\delta_{ef},\quad q_{f}^{(2)}=-I_{\mathrm{L}3}^{(f)}-\delta_{ef}.\label{eq:q1q2nue}
\end{equation}
Here $I_{\mathrm{L}3}^{(f)}$ is the third component of the weak isospin
of background fermions, $Q_{f}$ is the value of their electric charge,
$\theta_{\mathrm{W}}$ is the Weinberg angle, $\delta_{ef}=1$ for
electrons and vanishes for protons. The coefficients
$q_{f}^{(1,2)}$ in Eq.~(\ref{eq:q1q2nue}) correspond to the scattering
of electron neutrinos.

The plasma motion in an accretion disk around a rotating BH is quite
complex~\cite{AbrFra13}. A circular orbit is possible only when
an accretion disk is thin and is situated in the BH equatorial plane.
Since we study the general gravitational scattering of neutrinos,
which involves the neutrino motion both above and below the equatorial
plane (see Sec.~\ref{sec:PARAM} below), and consider the accretion
disk with a nonzero thickness, we assume that plasma in the accretion
disk is nonrelativistic and unpolarized. In this case, only $J_{f}^{0}=n_{f}U_{f}^{t}\neq0$,
whereas $J_{f}^{i}=0$ and $\Lambda_{f}^{\mu}=0$. Here $n_{f}$ is
the invariant number density of background fermions, measured by a
comoving observer, and $U_{f}^{t}$ is the time component of their
four velocity. We compute $U_{f}^{t}$ in~\ref{sec:UtDERIV};
see Eq.~(\ref{eq:Utnonrel}). Eventually, we get that the only nonzero
component of $G^{\mu}$ in Eq.~(\ref{eq:Gmu}) is $G^{t}=n_{e}U_{f}^{t}$,
where $n_{e}$ is the
electron number density and we suppose that plasma is electroneutral.

Using Eqs.~(\ref{eq:vierbKerr}) and~(\ref{eq:Utnonrel}), we obtain
that the vector $g^{a}=e_{\ \mu}^{a}G^{\mu}$ has the form,
\[
g^{a}=\frac{n_{e}U_{f}^{t}}{\sqrt{\Xi}}\left(\sqrt{\Sigma\Delta},0,0,-\frac{arr_{g}\sin\theta}{\sqrt{\Sigma}}\right).
\]
The distribution of matter in an accretion disk is quite model dependent.
We suppose that~\cite{SteSpr99}
\begin{equation}
n_{e}(r,\theta)=n_{e}^{(c)}\exp\left(-\frac{r^{2}\cos^{2}\theta}{2H^{2}}\right),\label{eq:nedistr}
\end{equation}
where $H$ is the disk thickness and $n_{e}^{(c)}$ is the central
density, i.e. the density at the equatorial plane. We take that~\cite{NarYi94}
$n_{e}^{(c)}(r)\propto n_{0}r^{-3/2}$, where $n_{0}=10^{18}\,\text{cm}^{-3}$
for SMBH with $M=10^{8}M_{\odot}$~\cite{Jia19}.

The matter contribution to the vector $\bm{\Omega}$, which determines
the spin precession of ultrarelativistic neutrinos in Eq.~(\ref{eq:vectG}),
takes the form,
\begin{equation}
\bm{\Omega}_{\mathrm{matt}}=\frac{G_{\mathrm{F}}}{\sqrt{2}U^{t}}\mathbf{u}\left(g^{0}-\frac{(\mathbf{gu})}{1+u^{0}}\right),\label{eq:Omegam}
\end{equation}
where we neglect the term $\propto\mathbf{g}/U^{t}$ for ultrarelativistic
neutrinos. Nevertheless both $\mathbf{u}/U^{t}$ and $\mathbf{u}/(1+u^{0})$
are finite in Eq.~(\ref{eq:Omegam}).

\section{Parameters of the system\label{sec:PARAM}}

We study the gravitational scattering of ultrarelativistic neutrinos
by a rotating SMBH surrounded with a magnetized accretion disk with
a finite thickness. The incoming flux of neutrinos is parallel to
the equatorial plane of BH. Nevertheless, unlike Refs.~\refcite{Dvo20a,Dvo20,Dvo21},
we do not restrict ourselves by the equatorial neutrino motion. Incoming
neutrinos are emitted from the direction $(\theta,\phi)_{s}=(\pi/2,0)$.
A remote neutrino detector is in the arbitrary position with the angular
coordinates $(\theta_{\mathrm{obs}},\phi_{\mathrm{obs}})$.

The mass of SMBH is taken to be $10^{8}M_{\odot}$. The accretion
disk around this SMBH consists of hydrogen plasma with the electron
density distribution in Eq.~(\ref{eq:nedistr}) (see also Refs.~\refcite{SteSpr99,Jia19}).
The main uncertainty among the disk parameters is its thickness $H$.
We vary it in the range $(1-10)r_{g}$ in our simulations (see, e.g.,
Ref.~\refcite{Sad09}), considering disks with a nonzero thickness.
We study the simplified model of the accretion disk which does not
account for the disk rotation. The neutrino interaction with plasma
is within the Fermi approximation of the standard model. We consider
only the forward scattering of electron neutrinos on plasma fermions.

We take into account the poloidal component of the magnetic field
which is based on the following vector potential in the world coordinates~\cite{Wal74}:
\begin{equation}
A_{t}=Ba\left[1-\frac{rr_{g}}{2\Sigma}(1+\cos^{2}\theta)\right],\quad A_{\phi}=-\frac{B}{2}\left[r^{2}+a^{2}-\frac{a^{2}rr_{g}}{\Sigma}(1+\cos^{2}\theta)\right]\sin^{2}\theta,\label{eq:Atphi}
\end{equation}
where the amplitude $B$ is supposed to scale with radius as $B\propto B_{0}r^{-5/4}$~\cite{BlaPay82}.
The strength $B_{0}$, which is the magnetic field at the inner radius of the disk, is taken to be $B_{0}=3.2\times10^{2}\,\text{G}$.
This value of $B_{0}$ is below the Eddington limit for $M=10^{8}M_{\odot}$~\cite{Bes10}.
The toroidal magnetic field, which is inevitably generated in a thick
disk and can be rather strong, is not considered in our model. We
plan to account for both the realistic plasma motion and the toroidal
magnetic field in one of the forthcoming works.

A neutrino is taken to be a Dirac particle with a nonzero magnetic
moment $\mu$. We consider the values of $\mu$ in the range $\mu=(10^{-14}-10^{-13})\mu_{\mathrm{B}}$.
Such magnetic moments are within the theoretical and astrophysical
constraints on neutrino magnetic moments established in Refs.~\refcite{Bel05,Via13}.

To reconstruct the neutrino motion, first, we find the $\theta(r)$
dependence. One can find the details of this procedure in Refs.~\refcite{Dvo23,Cha83}.
The knowledge of $\phi(r)$ is not necessary since the spin evolution
does not depend on $\phi$. We just need to obtain the final scattering
angle $\phi_{\mathrm{obs}}$. This fact significantly accelerates
numerical simulations allowing us to involve more test particles and
perform more frequent meshing in the radial direction. Now we use $2500$
test neutrinos in the incoming flux. However, some of these particles
are not involved in the scattering since they fall to the BH shadow
region.

The right hand side of Eq.~(\ref{eq:Schrod}) is given only in the
discrete nodes $r_{i}$. Moreover, the grid $r_{i}$ is irregular.
Thus, we apply the two-step Adams--Bashforth method for the numerical
integration of Eq.~(\ref{eq:Schrod}). The corresponding iterative
procedure is provided in~\ref{sec:ADAMS}; see Eqs.~(\ref{eq:Ad2})
and~(\ref{eq:C10}). It allows us to increase the accuracy of numerical
simulations compared to Ref.~\refcite{Dvo23} where the Euler method
was used.

\section{Results\label{sec:RES}}

In this section we present the results of the numerical solution of
Eq.~(\ref{eq:Schrod}) accounting for the values of the parameters
in Sec.~\ref{sec:PARAM}.

Ultrarelativistic neutrinos are emitted as left polarized particles
in frames of the standard model. A terrestrial neutrino telescope
can detect only left neutrinos as well. Thus, if $F_{0}\propto\mathrm{d}\sigma/\mathrm{d}\varOmega$
is the flux of scalar test particles, the observed flux of neutrinos
is $F_{\nu}=P_{\mathrm{LL}}F_{0}$. Here $\mathrm{d}\sigma/\mathrm{d}\varOmega$
is the differential cross section of the gravitational scattering.
Our main goal is to find the ratio $F_{\nu}/F_{0}$ for neutrinos
gravitationally scattered by BH and to account for the neutrino spin
precession in all external fields. We normalize the neutrino flux
by $F_{0}$ in order to avoid the singularities in the cross section
which are inherent in the gravitational scattering of both spinning
and spinless particles~\cite{BosWan74}. The detailed study of the
gravitational scattering of scalar test particles off a rotating BH
can be found in Ref.~\refcite{Boz08}. We utilize the term `scalar particles'
meaning their motion along null geodesics.

\begin{figure}
  \centering
  \subfigure[]
  {\label{fig:grava}
  \includegraphics[scale=.3]{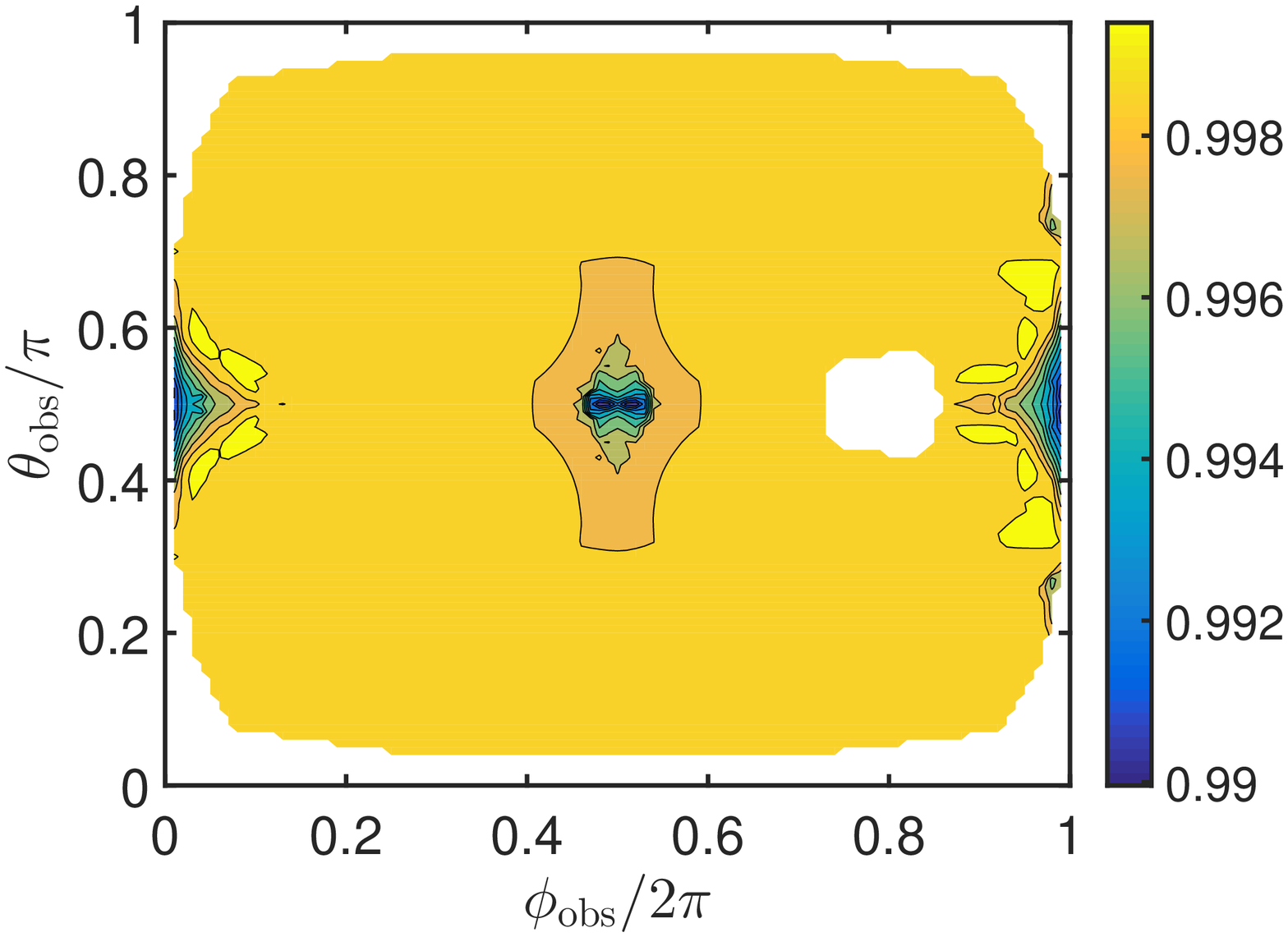}}
  \hskip-.3cm
  \subfigure[]
  {\label{fig:gravb}
  \includegraphics[scale=.3]{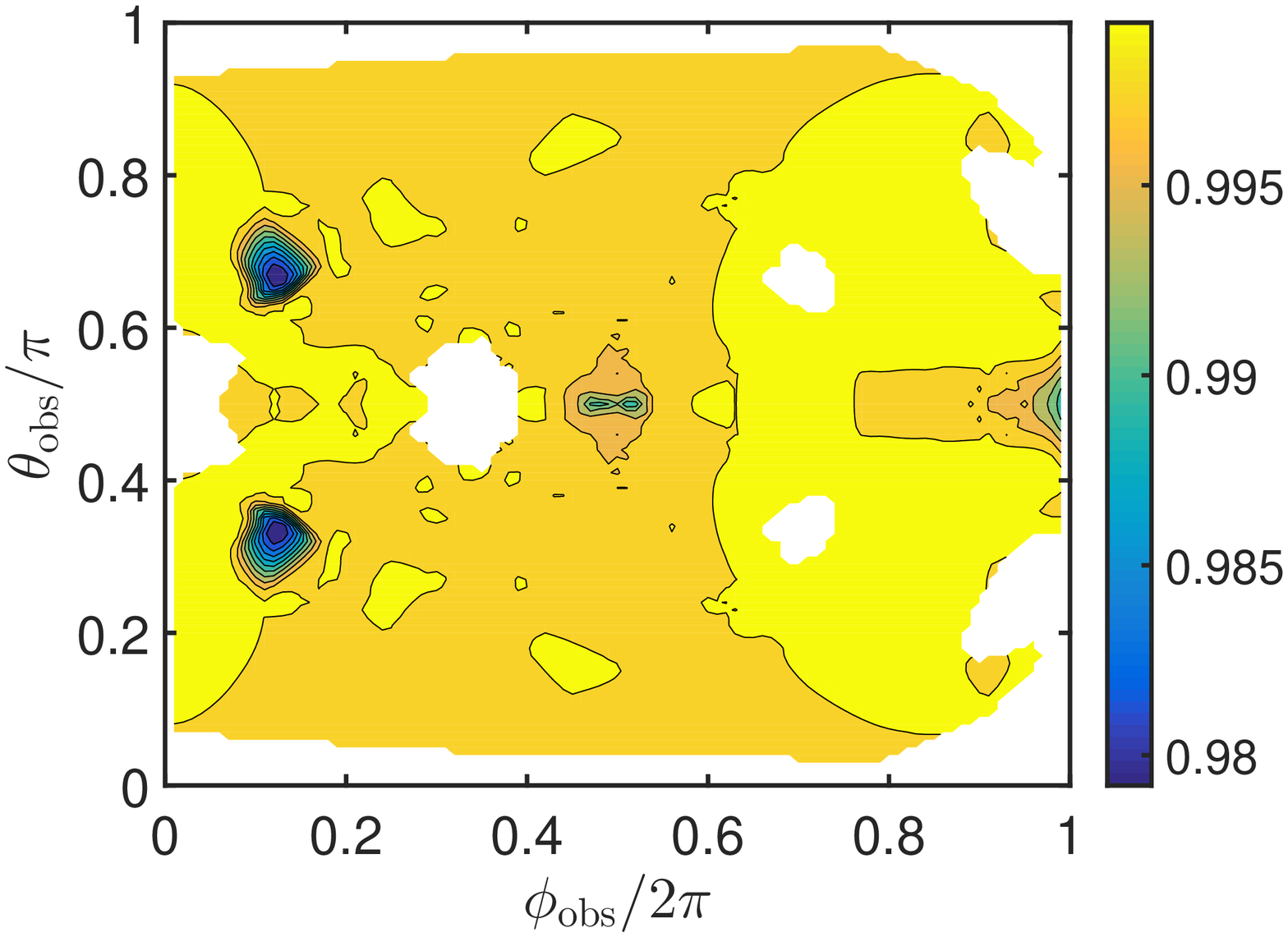}}
  \protect
\caption{The ratio of the fluxes of spinning ultrarelativistic neutrinos $F_{\nu}$
and the scalar particles $F_{0}$ scattered off BH with different
spins when only the gravitational interaction is accounted for. Panel
(a): $a=0.02M$; panel (b): $a=0.98M$.\label{fig:grav}}
\end{figure}

In the present work, we use the modified method for the integration
of Eq.~(\ref{eq:Schrod}); cf.~\ref{sec:ADAMS}. We should
check the behavior of the solution when only the gravitational interaction
is taken into account. We expect that $F_{\nu}/F_{0}\approx1$, i.e.
no spin oscillations occur in a purely gravitational scattering. The
corresponding general theorem has been proven in Ref.~\refcite{Dvo23}.
We show $F_{\nu}/F_{0}=P_{\mathrm{LL}}$ in Fig.~\ref{fig:grav}
for BHs with different angular momenta.

One can see in Fig.~\ref{fig:grav} that the deviation of $F_{\nu}/F_{0}$
from one is less than $1\%$ for the majority of $(\theta_{\mathrm{obs}},\phi_{\mathrm{obs}})$.
The higher accuracy of simulations compared to that in Ref.~\refcite{Dvo23}
is owing to the two-step Adams--Bashforth method represented in~\ref{sec:ADAMS}.
The result that $F_{\nu}/F_{0}\approx1$ means that our simulations
are reliable.

One has white areas Fig.~\ref{fig:grav} (see also Figs.~\ref{fig:a02}
and~\ref{fig:a098} below), which are present because of the insufficient
number of neutrinos scatter to these regions. Thus, 2D cubic interpolation,
used in our simulations, is unable to create contour plots there.
This shortcoming may be eliminated by a significant enhancement of
the test particles number. We plan to implement this task in a future
work.

Now, we add the neutrino interaction with the magnetic field and the
accretion disk to Eq.~(\ref{eq:Schrod}). First, we consider an almost
nonrotating SMBH with $a=0.02M$. We show the ratio $F_{\nu}/F_{0}$
in Fig.~\ref{fig:a02}. We plot $F_{\nu}/F_{0}$ in case when the
magnetic interaction is dominant, i.e. $V_{m}=0$, in Fig.~\ref{fig:a02a}
for $\mu=10^{-14}\mu_{\mathrm{B}}$ and in Fig.~\ref{fig:a02b}
for $\mu=10^{-13}\mu_{\mathrm{B}}$. In both cases, $B_{0}=3.2\times10^{2}\,\text{G}$.
Then, in Figs.~\ref{fig:a02c}-\ref{fig:a02f}, we account for
a nonzero interaction with background matter by setting the electron
number density near the SMBH surface to $n_{0}=10^{18}\,\text{cm}^{-3}$.
Figures~\ref{fig:a02c} and~\ref{fig:a02d} correspond to the
disk thickness $H=r_{g}$, whereas Figs.~\ref{fig:a02e} and~\ref{fig:a02f}
to $H=10r_{g}$.

\begin{figure}
  \centering
  \subfigure[]
    {\label{fig:a02a}
    \includegraphics[scale=.3]{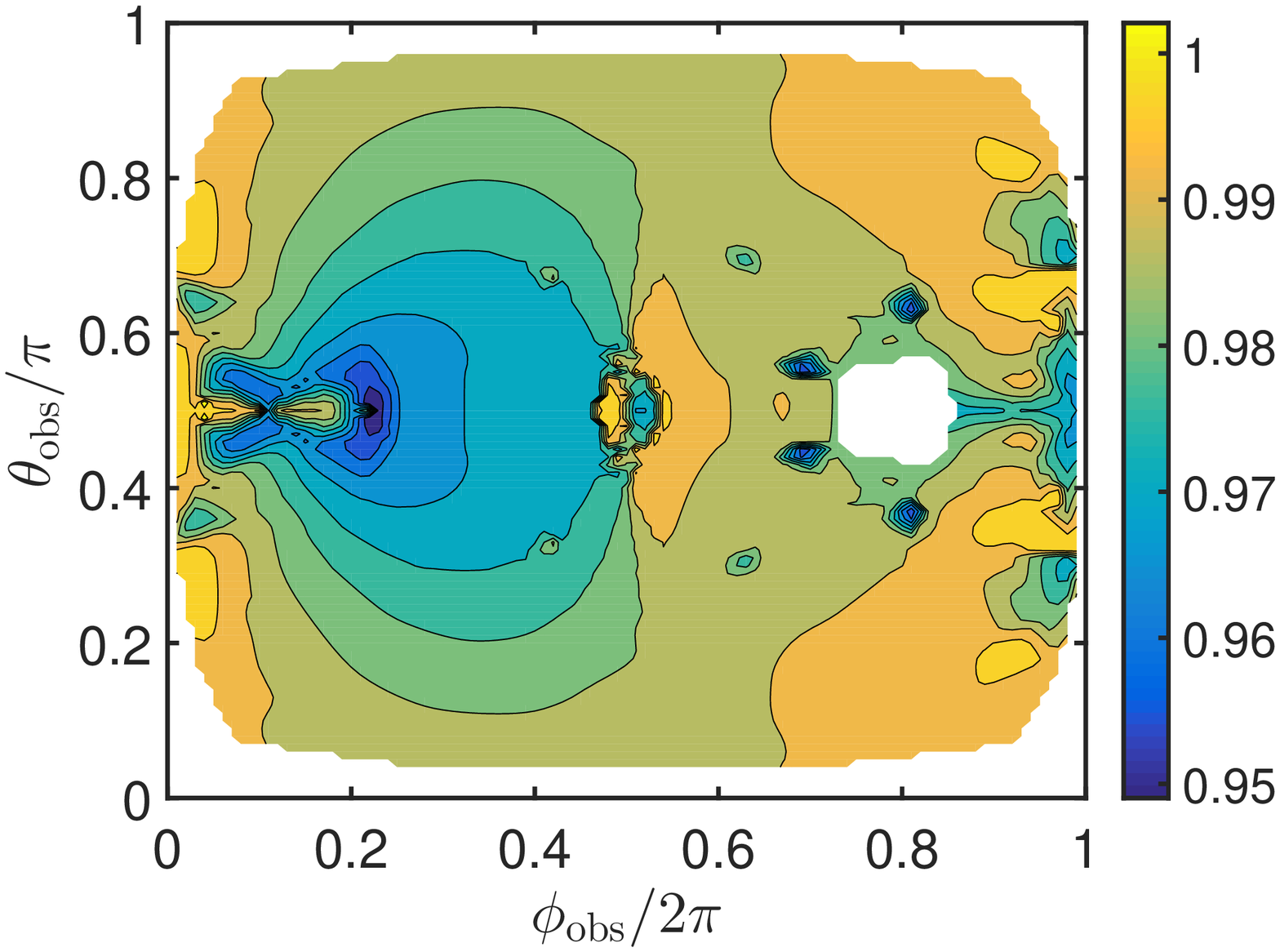}}
  \hskip-.3cm
  \subfigure[]
    {\label{fig:a02b}
    \includegraphics[scale=.3]{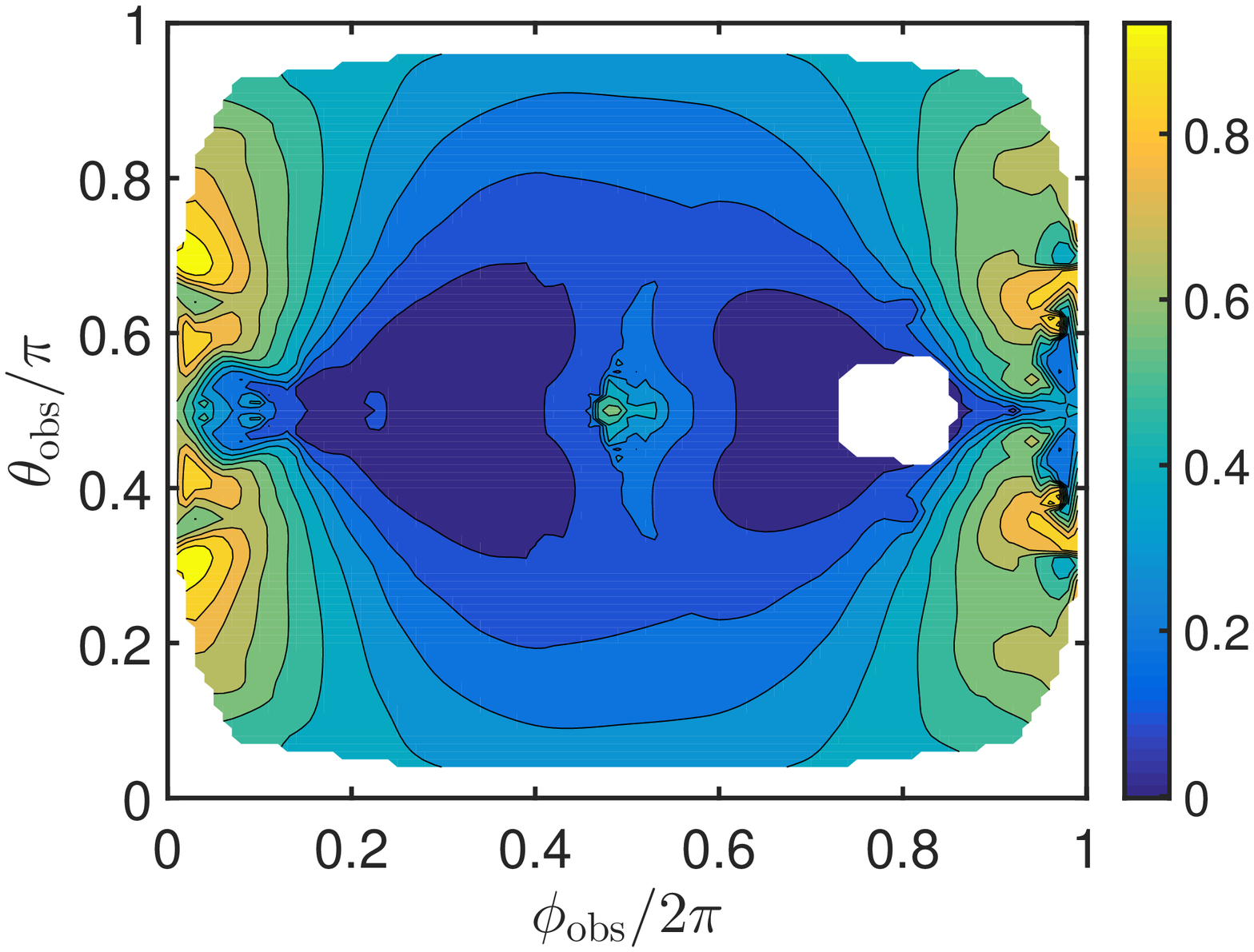}}
  \\
  \subfigure[]
    {\label{fig:a02c}
    \includegraphics[scale=.3]{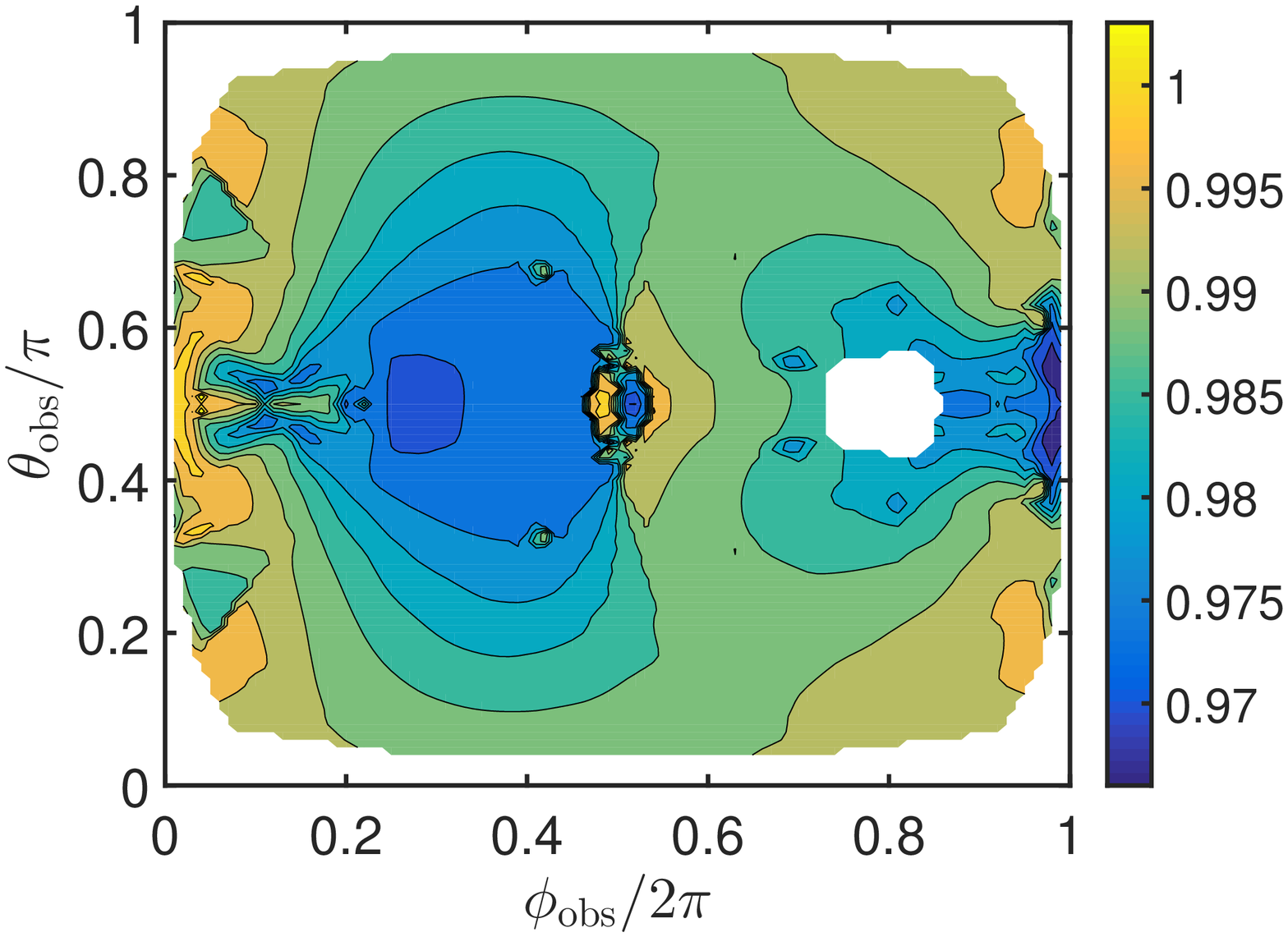}}
  \hskip-.3cm
  \subfigure[]
    {\label{fig:a02d}
    \includegraphics[scale=.3]{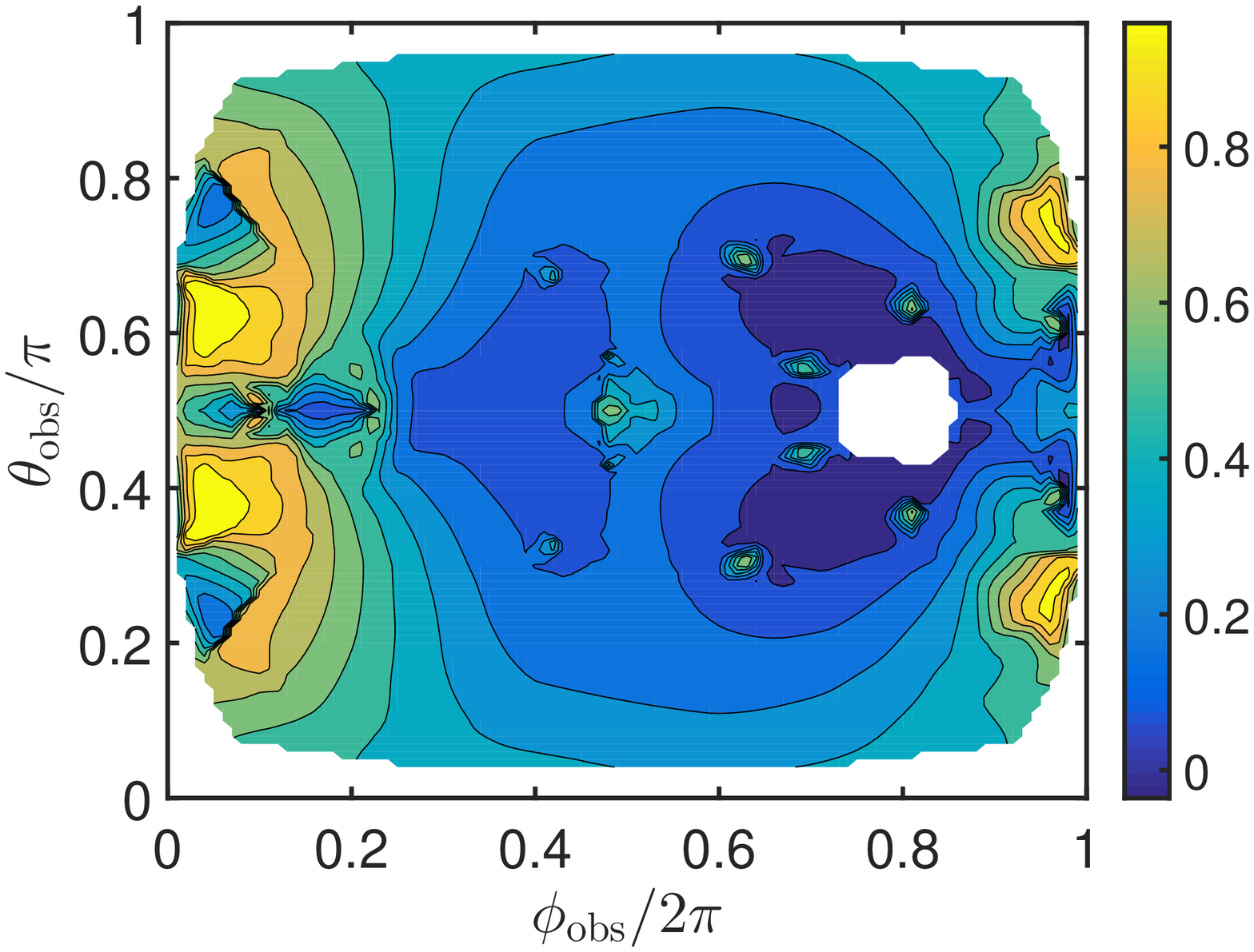}}
  \\
  \subfigure[]
    {\label{fig:a02e}
    \includegraphics[scale=.3]{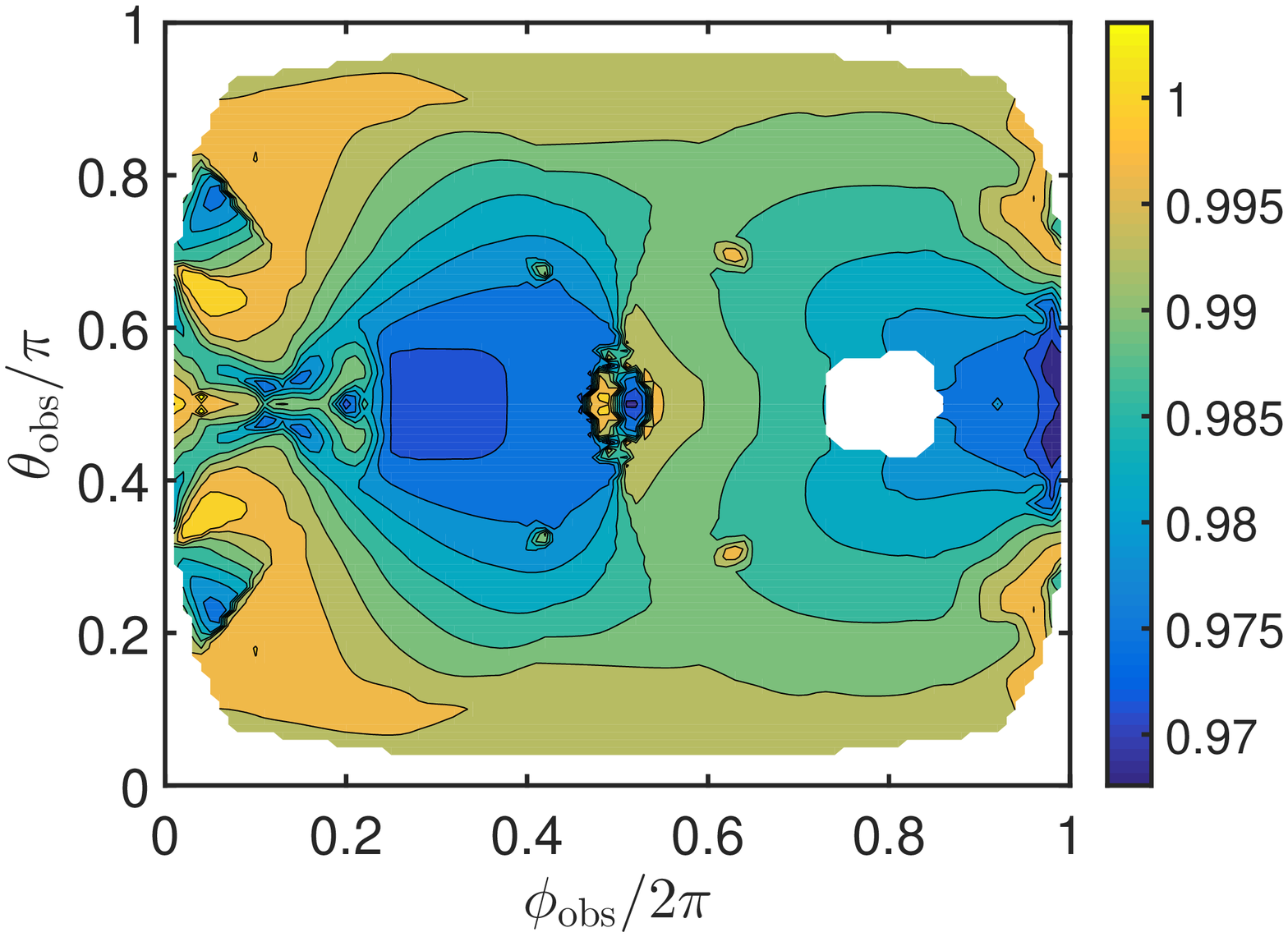}}
  \hskip-.3cm
  \subfigure[]
    {\label{fig:a02f}
    \includegraphics[scale=.3]{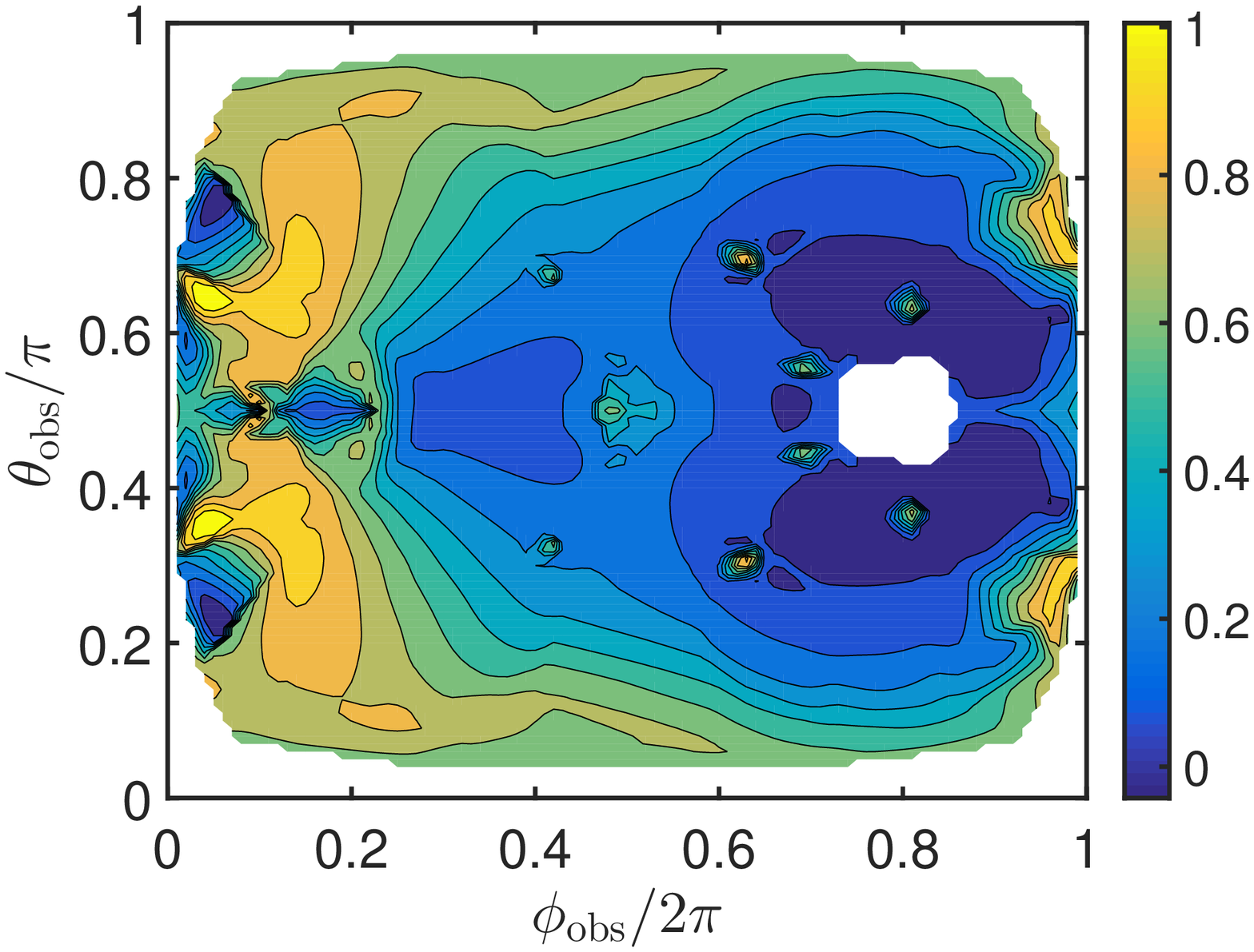}}
  \protect
\caption{The ratio of the fluxes $F_{\nu}/F_{0}$ for ultrarelativistic $\nu_{e}$
scattered off the almost nonrotating SMBH with $a=0.02M$ and $M=10^{8}M_{\odot}$.
Panel (a): $V_{\mathrm{B}}=\mu B_{0}r_{g}=2.7\times10^{-2}$, $V_{m}=G_{\mathrm{F}}n_{0}r_{g}/\sqrt{2}=0$.
Panel (b): $V_{\mathrm{B}}=2.7\times10^{-1}$, $V_{m}=0$. Panel (c):
$V_{\mathrm{B}}=2.7\times10^{-2}$, $V_{m}=10^{-1}$, $H=r_{g}$.
Panel (d): $V_{\mathrm{B}}=2.7\times10^{-1}$, $V_{m}=10^{-1}$, $H=r_{g}$.
Panel (e): $V_{\mathrm{B}}=2.7\times10^{-2}$, $V_{m}=10^{-1}$, $H=10r_{g}$.
Panel (f): $V_{\mathrm{B}}=2.7\times10^{-1}$, $V_{m}=10^{-1}$, $H=10r_{g}$.\label{fig:a02}}
\end{figure}

We can see in Figs.~\ref{fig:a02c} and~\ref{fig:a02d} that
the neutrino interaction with plasma of the accretion disk suppresses
neutrino spin oscillations. We remind that spin oscillations of neutrinos
are in the resonance when only the magnetic interaction is taken into
account (see, e.g., Ref.~\refcite{VolVysOku86}, where this problem is
discussed in the flat spacetime). The electroweak interaction with
matter shifts spin oscillations out of the resonance. It explains
the behavior of $P_{\mathrm{LL}}$ in Figs.~\ref{fig:a02c} and~\ref{fig:a02d}.
Moreover, the consideration of a thicker disk in Figs.~\ref{fig:a02e}
and~\ref{fig:a02f} results in the further enhancement of the survival
probability since the regions with the great $P_{\mathrm{LL}}$ become
wider. This result is explained by the fact that more scattered neutrinos
interact with matter if the disk is thicker.

Finally, we plot $F_{\nu}/F_{0}$ in Fig.~\ref{fig:a098} for an
almost maximally rotating SMBH with $a=0.98M$. Qualitatively, the
behavior of the survival probability resembles that in Fig.~\ref{fig:a02},
which is discussed above. We just mention that the interaction with
matter is more pronounced for a small neutrino magnetic moment. Indeed,
one can see in Figs.~\ref{fig:a098c} and~\ref{fig:a098e} that
neutrino spin oscillations are almost suppressed.

\begin{figure}
  \centering
  \subfigure[]
    {\label{fig:a098a}
    \includegraphics[scale=.3]{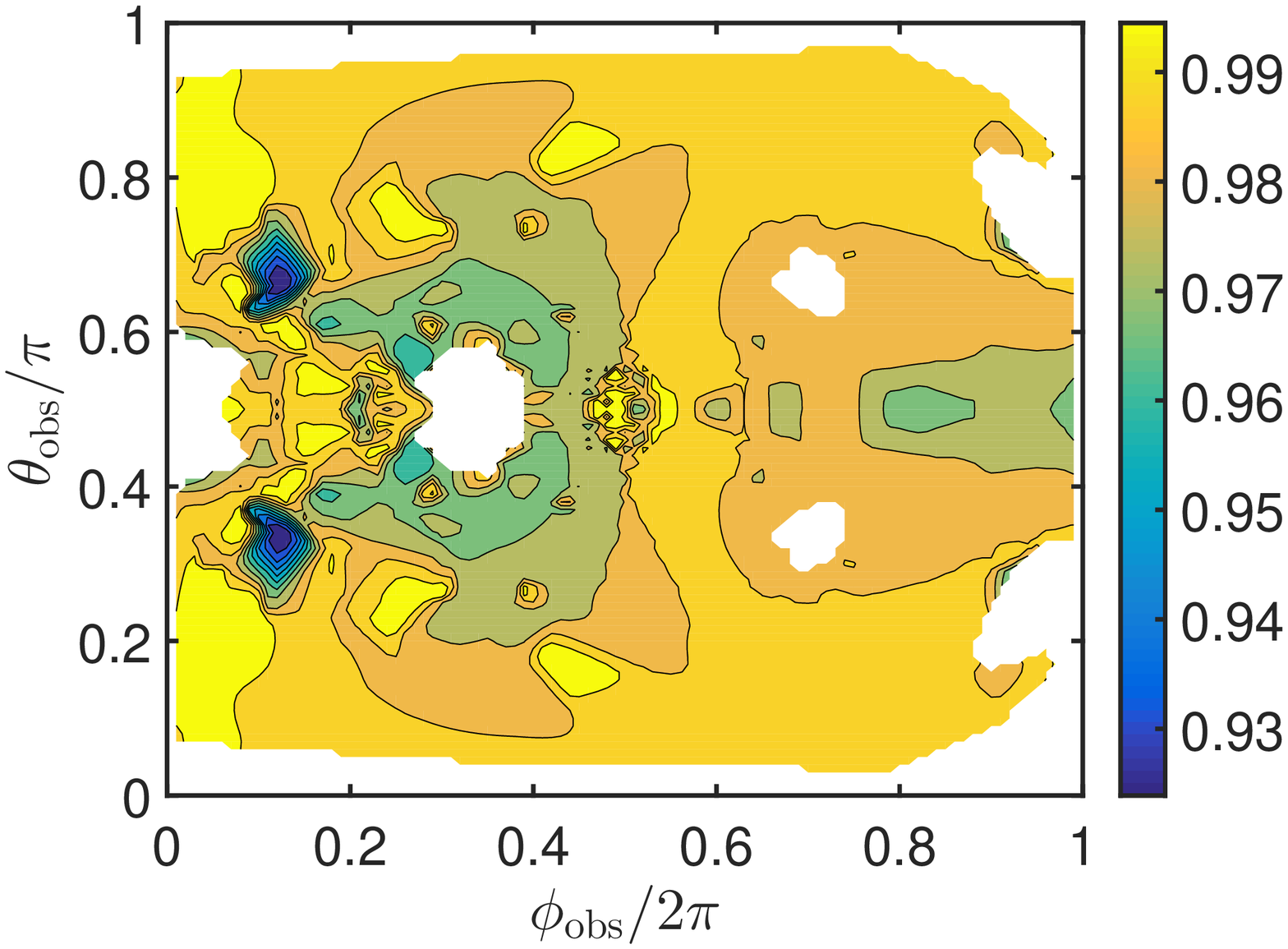}}
  \hskip-.3cm
  \subfigure[]
    {\label{fig:a098b}
    \includegraphics[scale=.3]{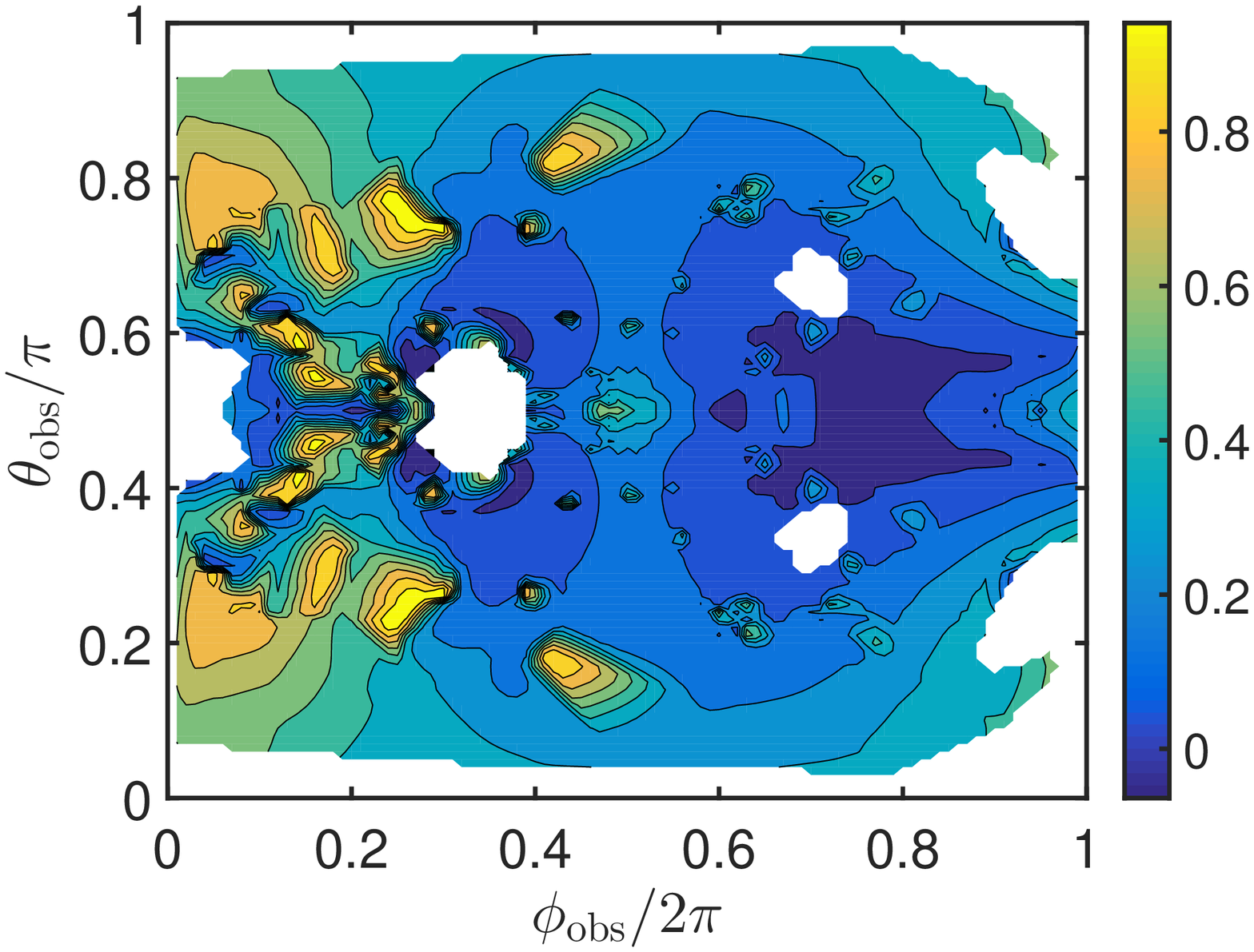}}
  \\
  \subfigure[]
    {\label{fig:a098c}
    \includegraphics[scale=.3]{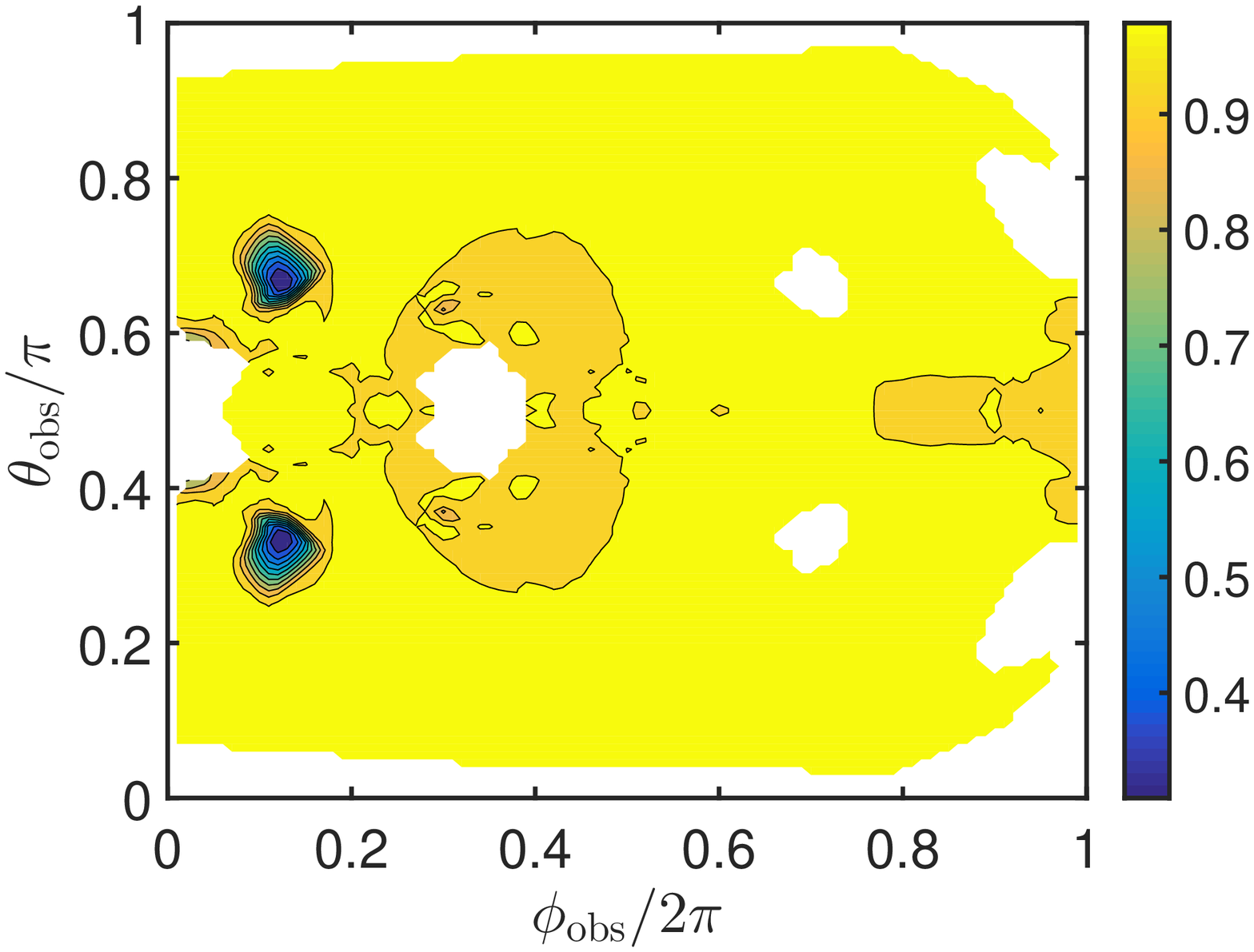}}
  \hskip-.3cm
  \subfigure[]
    {\label{fig:a098d}
    \includegraphics[scale=.3]{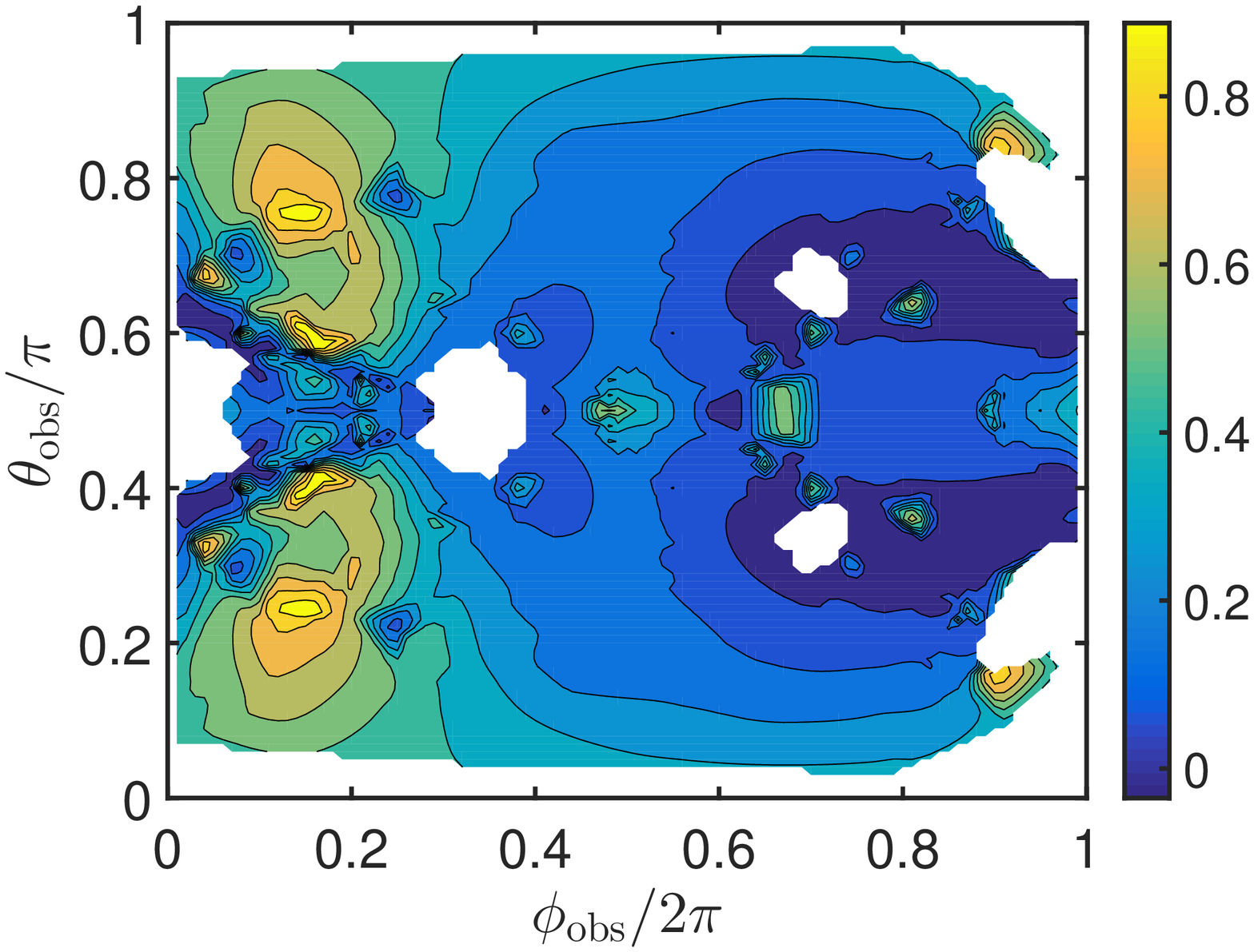}}
  \\
  \subfigure[]
    {\label{fig:a098e}
    \includegraphics[scale=.3]{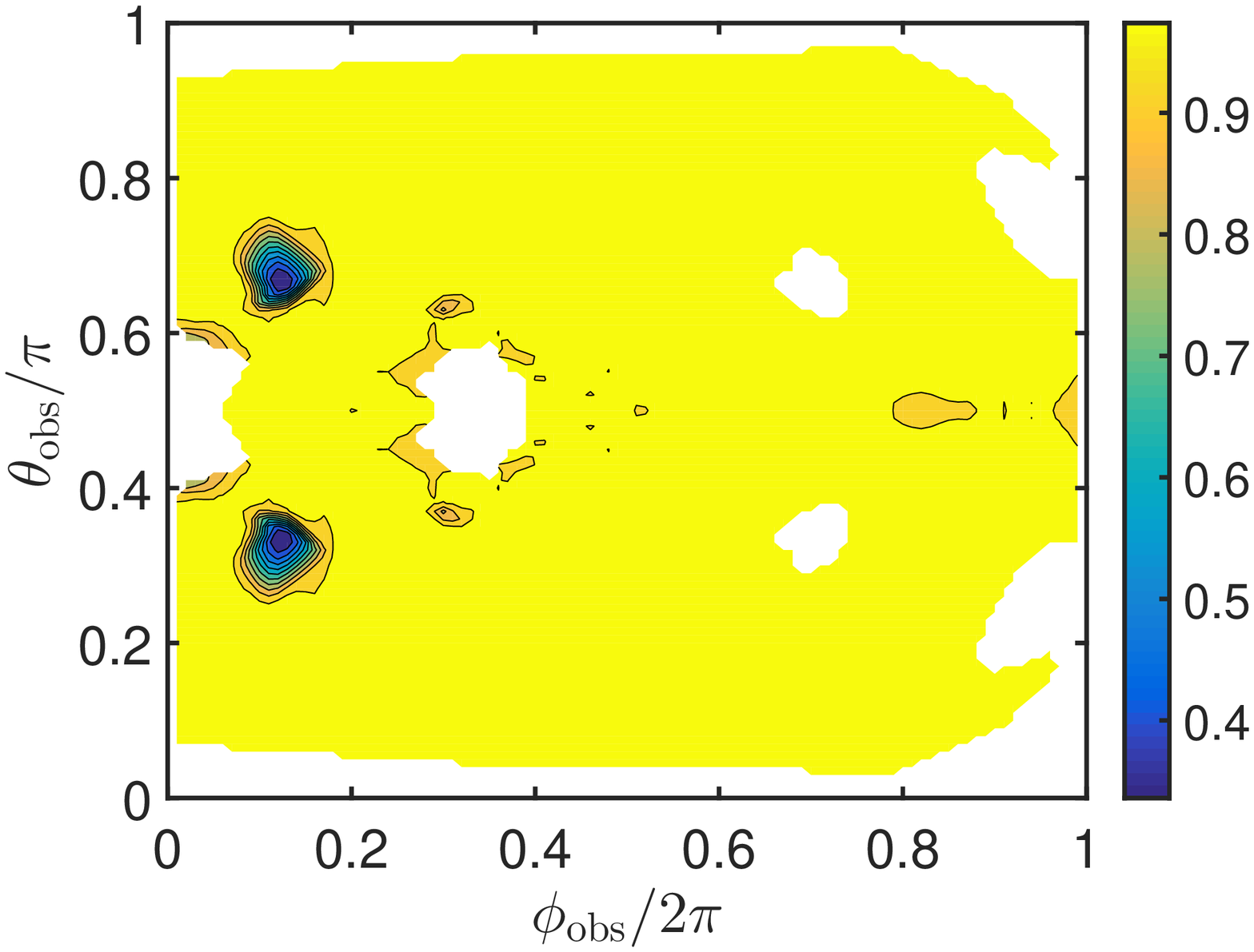}}
  \hskip-.3cm
  \subfigure[]
    {\label{fig:a098f}
    \includegraphics[scale=.3]{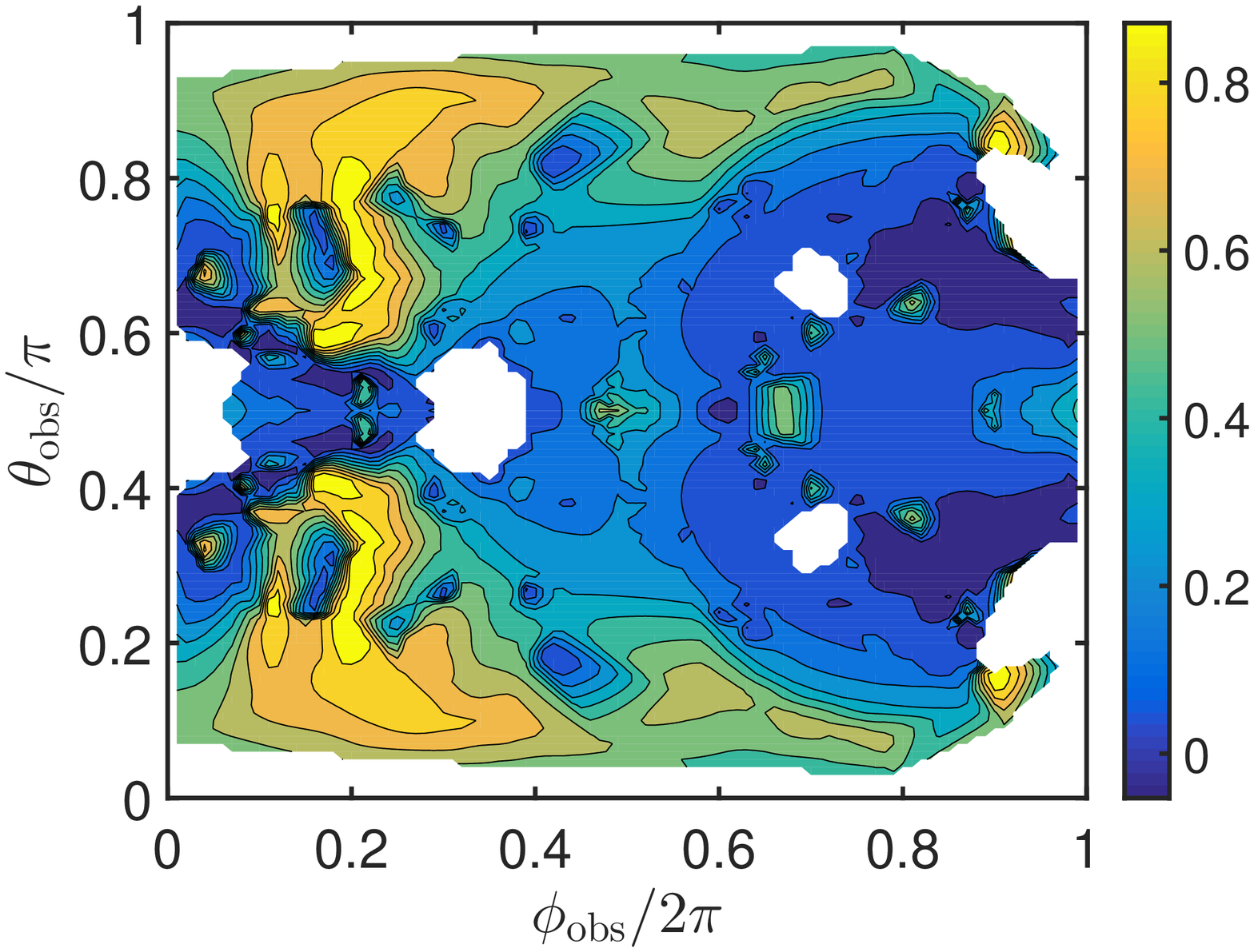}}
  \protect
\caption{The same as in Fig.~\ref{fig:a02} for $a=0.98M$, i.e. SMBH is almost
maximally rotating.\label{fig:a098}}
\end{figure}

\section{Conclusion\label{sec:CONCL}}

In the present work, we have studied the scattering of ultrarelativistic
neutrinos off a rotating SMBH surrounded by a magnetized accretion
disk with a nonzero thickness. The flux of left polarized incoming
neutrinos was taken to propagate parallel to the equatorial plane
of BH. However, we did not restrict ourselves to a purely equatorial
neutrino motion. These neutrinos have been strongly gravitationally
lensed towards a distant observer with the arbitrary angular coordinates
$0<\theta_{\mathrm{obs}}<\pi$ and $0<\phi_{\mathrm{obs}}<2\pi$.

Unlike scalar particles or photons, the polarization of neutrinos
is important for their detection since we cannot observe sterile right
neutrinos. That is why, besides the reconstruction of the particle
trajectories in the curved spacetime, we had to study neutrino spin
precession in the external fields. We have supposed that a neutrino
is a Dirac particle having a nonzero magnetic moment, which provides
the interaction with a poloidal magnetic field in the accretion disk.
The neutrino electroweak interaction with plasma in the disk is considered
in frames of the Fermi approximation.

In Sec.~\ref{sec:MOTION}, we have outlined the neutrino motion in
the Kerr metric and the description of the spin evolution in external
fields in curved spacetime. The neutrino interaction with background
matter and the model of the accretion disk have been described in
Sec.~\ref{sec:MATTER}. The characteristics of the magnetic field
are the same as in Ref.~\refcite{Dvo23}. Therefore we did not discuss
them in details here. Then, in Sec.~\ref{sec:PARAM}, we have specified
the rest of the parameters in the system and considered some details
of the numerical simulations.

We have presented our results in Sec.~\ref{sec:RES}. The fluxes
of spinning neutrinos have been found for various parameters of the
system. We have confirmed the general theorem, proven in Ref.~\refcite{Dvo23},
that the polarization of ultrarelativistic neutrinos is unchanged
in a purely gravitational scattering. Then, we have obtained that
the neutrino interaction with plasma of an accretion disk suppresses
neutrino spin oscillations. This feature, known in flat spacetime,
has been demonstrated for the neutrino gravitational scattering. Besides
the consideration of the accretion disk with a finite thickness, one
of the advantages of the present work, compared to Ref.~\refcite{Dvo23},
was the more precise numerical integration of Eq.~(\ref{eq:Schrod}).
It allowed us to significantly increase the accuracy of the simulations.

The derivation of the time component of the four velocity of matter in the accretion disk has been provided in~\ref{sec:UtDERIV}.
In~\ref{sec:ADAMS}, we have described the more precise method
of the integration of Eq.~(\ref{eq:Schrod}) in a nonuniform grid.

The results obtained allow one to explore the characteristics of inner
parts of the accretion disk around SMBH in our Galaxy, e.g., with
neutrinos emitted in a SN explosion. The huge number of such neutrinos
are expected to be observed with the existing or future neutrino telescopes.
If such neutrinos are gravitationally lensed by the central SMBH,
we have to account for the interaction with strong external fields
in curved spacetime near this SMBH.

\appendix

\section{Derivation of $U^{t}$ for an arbitrary position of a plasma particle\label{sec:UtDERIV}}

The action for a test particle moving in the Kerr metric has the form,
$S=-Et+L\phi+\dotsc,$ where we show only the terms responsible for
the conserved energy, $E$, and the projection of the particle angular momentum
on the BH spin, $L$. The canonical four momentum of a massive particle
reads $p_{\mu}=mU_{\mu}=-\partial_{\mu}S$. Thus, we have that
\begin{equation}\label{eq:E-L}
  m
  \left(
    g_{tt}U^{t}+g_{t\phi}U^{\phi}
  \right) = E,
  \quad
  m
  \left(
    g_{\phi t}U^{t}+g_{\phi\phi}U^{\phi}
  \right)=-L.
\end{equation}
Solving Eq.~(\ref{eq:E-L}), we get that~\cite{Wil12}
\begin{equation}\label{eq:Utgen}
  U^{t}=-\frac{(Eg_{\phi\phi}+Lg_{\phi t})}{m(g_{\phi t}^{2}-g_{\phi\phi}g_{tt})}.
\end{equation}
Using the Kerr metric components in Eq.~(\ref{eq:Kerrmetr}), one
rewrites Eq.~(\ref{eq:Utgen}) in the form,
\begin{equation}\label{eq:Utexpl}
  U^{t}=\frac{E
  \left[
    \left(
      r^{2}+a^{2}
    \right)
    \Sigma+rr_{g}a^{2}\sin^{2}\theta
  \right]-
  Lrr_{g}a\sin^{2}\theta}{m\Sigma\Delta\sin^{2}\theta}.
\end{equation}
If a particle moves in the equatorial plane with $\theta=\pi/2$,
we reproduce the expression for $U^{t}$ obtained in Ref.~\refcite{Dvo20}.

Finally, if a particle is nonrelativistic, it has $E=m$ and $L=0$.
We obtain, basing on Eq.~(\ref{eq:Utexpl}), that
\begin{equation}\label{eq:Utnonrel}
  U^{t}=\frac{\Xi}{\Sigma\Delta\sin^{2}\theta},
\end{equation}
which is used in Sec.~\ref{sec:MATTER} as $U_{f}^{t}$.

\section{Two-step Adams--Bashforth method for an irregular grid\label{sec:ADAMS}}

We discuss the solution of the single first order differential equation
\begin{equation}
y'=f(x,y).\label{eq:eqtosol}
\end{equation}
The generalization of the results to a system of equations is straightforward.
If the function $f(x,y)$ in Eq.~(\ref{eq:eqtosol}) is given only
in discrete points $f_{i}$, we cannot use the precise Runge-Kutta
method.

The value of the function at the $(i+2)$th node is
\begin{equation}\label{eq:yi+1}
  y_{i+2}=y_{i+1}+\int_{x_{i+1}}^{x_{i+2}}f(x,y(x))\mathrm{d}x.
\end{equation}
We replace the integrand in Eq.~(\ref{eq:yi+1}) with the linear
polynomial
\begin{equation}\label{eq:fp}
  f(x,y(x))\to p(x)=\frac{x-x_{i}}{x_{i+1}-x_{i}}f_{i+1}+\frac{x-x_{i+1}}{x_{i}-x_{i+1}}f_{i}.
\end{equation}
One can see that $p(x_{i})=f_{i}$ and $p(x_{i+1})=f_{i+1}$.

Integrating in Eq.~(\ref{eq:yi+1}) and accounting for Eq.~(\ref{eq:fp}),
we get that
\begin{equation}\label{eq:Ad2}
  y_{i+2}=y_{i+1}+C_{1}f_{i+1}+C_{0}f_{i},
\end{equation}
where
\begin{equation}\label{eq:C10}
  C_{1}=\frac{(x_{i+2}-x_{i+1})}{2(x_{i+1}-x_{i})}(x_{i+2}+x_{i+1}-2x_{i}),
  \quad
  C_{0}=\frac{(x_{i+2}-x_{i+1})}{2(x_{i+1}-x_{i})}(x_{i+1}-x_{i+2}).
\end{equation}
If the grid is regular with the step $h$, we obtain in Eq.~(\ref{eq:C10})
that $C_{1}=\tfrac{3}{2}h$ and $C_{0}=-\tfrac{1}{2}h$. It reproduces
the result known previously (see, e.g., Ref.~\refcite{KorKor68}).

Equations~(\ref{eq:Ad2}) and~(\ref{eq:C10}) are more precise than
the Euler method. We use them in Sec.~\ref{sec:RES} to solve Eq.~(\ref{eq:Schrod}),
where the grid is irregular.

\end{document}